\documentclass
[aps,twocolumn,prl,superscriptaddress,showpacs,tightenlines]{revtex4}%
\usepackage{amssymb}
\usepackage{amsmath}
\usepackage{graphicx}
\usepackage{epsfig}
\usepackage{amsfonts}%
\providecommand{\U}[1]{\protect\rule{.1in}{.1in}}
\begin{document}
\title{Cooling Torsional Nanomechanical Vibration by Spin-Orbit Interactions}
\author{Nan Zhao}
\affiliation{Department of Physics, Tsinghua University, Beijing
100084, China}
\author{D.L. Zhou}
\affiliation{Institute of Physics, Chinese Academy of Science, Beijing 100080, China}
\author{Jia-Lin Zhu}
\email{zjl-dmp@tsinghua.edu.cn} \affiliation{Department of Physics,
Tsinghua University, Beijing 100084, China}
\author{C.P. Sun}
\email{suncp@itp.ac.cn, http://www.itp.ac.cn/~suncp}
\affiliation{Institute of Theoretical Physics, Chinese Academy of Science, Beijing 100080, China}
\date{\today}

\begin{abstract}
We propose and study a spin-orbit interaction based mechanism to
actively cool down the torsional vibration of a nanomechanical
resonator made by semiconductor materials. We show that the
spin-orbit interactions of electrons can induce a coherent coupling
between the electron spins and the torsional modes of nanomechanical
vibration. This coherent coupling leads to an active cooling for the
torsional modes via the dynamical thermalization of the resonator
and the spin ensemble.

\end{abstract}

\maketitle

%\qquad\label{I1}\label{I2}\label{I3}

\emph{Introduction}.-Fast developments of nano-fabrication
technology enable us to manufacture the nanomechanical resonators
(NAMRs) with high frequency and high
quality\cite{NEMSreviewPhysToday2005}. These artificial systems have
attracted more and more attentions for their application potentials,
such as ultra-sensitive probe for tiny
displacement\cite{NatureNEMSdisplacementSense2003}, single spin
detector\cite{NatureNEMSSpinDetection2004}, and physical
implementations of quantum
information\cite{PRLNEMSentanglement2004}. On the other hand, it is
believed that a high frequency NAMR (e.g. with GHz oscillation) will
exhibit various quantum effects in this macroscopic system at low
temperatures\cite{NEMSreviewPhysToday2005}. Therefore, the studies
on NAMR are relevant to fundamental problems like quantum
measurement\cite{bookQuantumMeasurementBraginsky}.

To give prominence to the quantization effects of NAMR, the crucial
issue depends on whether we can cool it down to the vibrational
ground state. Actually, interacting with the surrounding
environment, the NAMR reaches a thermal equilibrium through the
relaxation. The mean thermal occupation number of the NAMR, which is
determined by the environment temperature and the vibration
frequency, is always larger than unity for most of the present
nanomechanical systems even at the sub-Kelvin temperature.
Accordingly, besides decreasing the environment temperature, special
cooling mechanism is needed to be invented to further reduce the
thermal occupation number.

Recent experiments have shown that the micromechanical systems can
be cooled down by radiation pressure\cite{NatureNanoBeamCooling2006}
or by coupling to a single electron
transistor\cite{ScienceSQLofNAMR2004}. And several theoretical
scenarios were proposed based on different cooling mechanism, such
as laser excitation of phonon sideband of a quantum
dot\cite{PRLCoolingNAMRZoler2005}, and periodic coupling to a Cooper
pair box\cite{PRLCoolingNAMRSunCPZhangP2005}, et al. All the cooling
mechanisms mentioned above are only focused on the flexural
mechanical modes. In this Letter, we propose an active cooling
mechanism for the quantized torsional nanomechanical modes
(TNMMs)\cite{{PRATorsionalMechanics1999},{PRBDissipationNAMR2002},{NatureNanoTorsionCNT2006}}
of the NAMR.
\begin{figure}[pb]
\includegraphics[bb=5 10 485 325,angle=0, width=6.5 cm, clip]{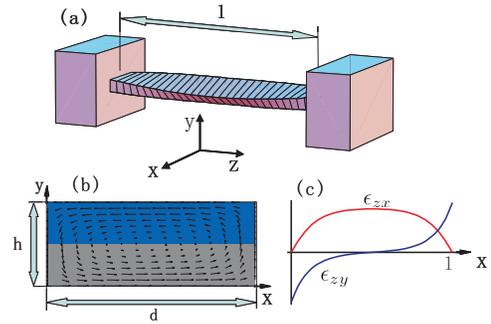}\caption{(a)
Schematic illustration of torsional vibration of the nano-mechanical
rod. (b) The cross section of the nanomechanical rod and its strain
distribution. Electrons flowing through the nanomechanical rod are
confined in the semiconductor layer (blue), which grows on an
insulating substrate (grey). (c) A typical strain distribution of
$\epsilon_{z x}$ (red) and $\epsilon_{z y}$ (blue). It is shown that
$\epsilon_{z x}$ is an even function with respect to
$x$, while\ \ $\epsilon_{z y}$ is odd.}%
\label{XRef-Figure-41095545}%
\end{figure}

Similar to the displacements and momentums for the flexural modes,
the TNMMs are described by the oscillation of the torsion angles and
the angular momentums of the NAMR. Thus, it is quite natural to to
consider how to couple the TNMMs to the electron spin degrees of
freedom, which also behave as angular momentums. Recently, a novel
interaction between the spin current in semiconductor materials and
the NAMR induced by the spin-orbit coupling is discovered in Ref.
\cite{PRLStrainSpinCoupling2005}. This pioneer investigation is
carried out in the semi-classical regime to provide the possibility
for detecting spin current by the
NAMR\cite{{PRLStrainSpinCoupling2005},{PRBMechanicalandSpinCurrent2004}}.

Here, we present a microscopic quantum mechanical description for
the coherent coupling between the electron spins and the TNMM. With
this coherent coupling, we show that the mean occupation number of
the NAMR can be significantly reduced. Thus, the NAMR is actively
cooled down and can be brought to the quantum realm.

\emph{Model for spin-strain coupling}.-We consider a doubly clamped
nanomechanical rod with length $l$, width $d$, and hight $h$
(Fig.\ref{XRef-Figure-41095545}). Similar to the structure
considered in Ref. \cite{PRLStrainSpinCoupling2005}, the
semiconductor layer grows on an insulating substrate (Fig.
\ref{XRef-Figure-41095545}(b)). The electrons flowing through the
nanomechanical rod are confined in the semiconductor layer. For the
$n$th normal mode, the torsion angle of the NAMR is
$\theta_{n}(z,t)=\theta_{n0}\sin(n\pi z/l)\cos(\omega_{n}t)$ with
frequency $\omega_{n}$ and amplitude $\theta_{n0}$.

According to the isotropic elastic theory\cite{{bookElasticLandau,
bookFoundationsofNEMSCleland}}, the strain tensor field has
non-vanishing components
$\epsilon_{xz}=\epsilon_{zx}=\theta^{\prime}(z)(\partial f/\partial
y)$ and $\epsilon_{yz}=\epsilon_{zy}=-\theta^{\prime}(z)(\partial
f/\partial x)$ where $\theta^{\prime}(z)=\partial\theta/\partial z$,
and the function $f(x,y)$ is determined by the cross section
geometry of the NAMR\cite{bookElasticLandau}. We illustrate the
strain field explicitly in Fig. \ref{XRef-Figure-41095545}(b).

To describe the quantum phenomena of the torsional oscillation of
the NAMR, we apply the standard canonical quantization procedure by
modeling the NAMR as
the many-harmonic-oscillator with Hamiltonian $H_{\operatorname{NAMR}}%
=\sum_{n}\hbar\omega_{n}(a_{n}^{\dagger}a_{n}+1/2)$. Here, the boson operator
$a_{n}=I^{1/4}\sqrt{M\omega_{n}/2\hbar}(\theta_{n0}+i\dot{\theta}_{n0}%
/\omega_{n})$ is introduced through the effective mass
$M=\rho\sqrt{I}l/2$, where $\rho$ is the mass density, and
$I=\int(x^{2}+y^{2})dA$ is the rotating inertia about $z$ axis for
$A$ being the cross section area.

In the following, we will focus on the fundamental ($n=1$) TNMM at
low temperatures and thus omit the subscript index of the boson
operators. Then the \emph{quantized} strain field along the NAMR is
described by $\epsilon _{zx}=(\partial f/\partial
y)F(z),\epsilon_{zy}=-(\partial f/\partial x)F(z)$, where
$F(z)=\epsilon_{1}\cos(qz)(a^{\dagger}+a)$, $\epsilon_{1}=\pi
l^{-1}(\hbar/2M\omega_{1}\sqrt{I})^{1/2}$ and $q=\pi/l$.

As pointed out in Ref. \cite{PRLStrainSpinCoupling2005}, the spins
in a nanomechanical rod can be coupled to the strain field in narrow
band\ \ semiconductor materials by the
Hamiltonian\cite{bookOpticalOrientation}
\begin{equation}
H_{\operatorname{SO}}^{3D}=\alpha\lbrack\sigma_{x}(\epsilon_{xz}k_{z}%
-\epsilon_{xy}k_{y})+c.p.],
\end{equation}
where $\alpha$ is the coupling strength, and $\sigma_{x,y,z}$ and
$\hbar k_{x,y,z}$ are the Pauli matrices and the momentum components
of the electron respectively. Here, `c.p.' stands for cyclic
permutation of $x$, $y$ and $z$. Due to the confinement in the $x$
and $y$ directions, the spacial wave function of the electron can be
separated as $\psi(x,y,z)=\varphi (x,y)\exp(ik_{z}z)$, where
$\varphi(x,y)$ is determined by the transverse confinement in the
$x$ and $y$ directions. We further assume that this confinement is
so strong that we can take an average over $\varphi(x,y)$ to obtain
an effective Hamiltonian $H_{\operatorname{SO}}=\langle\varphi
(x,y)|H_{\operatorname{SO}}^{3D}|\varphi(x,y)\rangle$. Considering
the parity of the non-vanishing strain components (see Fig.
\ref{XRef-Figure-41095545}(c)), the Hamiltonian
$H_{\operatorname{SO}}^{3D}$ is then reduced from three-dimension to
one-dimension effectively, i.e. $H_{\operatorname{SO}}=\alpha{\mathcal{E}%
}k_{z}(a^{\dagger}+a)\cos(qz)\sigma_{x}$, where ${\mathcal{E}}=\langle\partial
f/\partial y\rangle\epsilon_{1}$ is a dimensionless constant, and
$\langle\partial f/\partial y\rangle$ denotes the average over the wave
function $\varphi(x,y)$.

Next, we further simplify the single electron Hamiltonian by
considering the classical limit of the spatial motion of the
electron. In the case of large longitudinal momentum of the injected
electron compared with the wave number of the TNMM, i.e. $k_{z}\ \
>>q$, the longitudinal motion of the injected electron is hardly
affected by the back action of the vibration. Thus, we neglect the
recoil effect and used the approximate Galilean transformation
$z\rightarrow z_{0}+vt$, where $v=\hbar k_{z}/m$ is the velocity of
the coherent injected electron. With this approximation, the
Hamiltonian can be written as:
\begin{equation}
H=\frac{1}{2}\hbar\omega_{z}\sigma_{z}+\hbar\omega_{1}a^{\dagger}a+g_{v}%
\cos(\omega_{v}t+\phi)\left(  a^{\dagger}+a\right)  \sigma_{x}%
,\label{XRef-Equation-11522235}%
\end{equation}

\noindent where $\hbar\omega_{z}$ is the Zeeman energy of the electron spin.
We have ignored the constant term of kinetic energy $\hbar^{2}k_{z}^{2}/2m$.
Here, we define the velocity dependent coupling constant $g_{v}=\alpha
{\mathcal{E}} m v/\hbar$, the frequency $\omega_{v}=q v$, and the phase
$\phi=q z_{0}$.

\begin{figure}[ptb]
\includegraphics[bb=40 54 555 810,angle=-90, width=7.2 cm, clip]{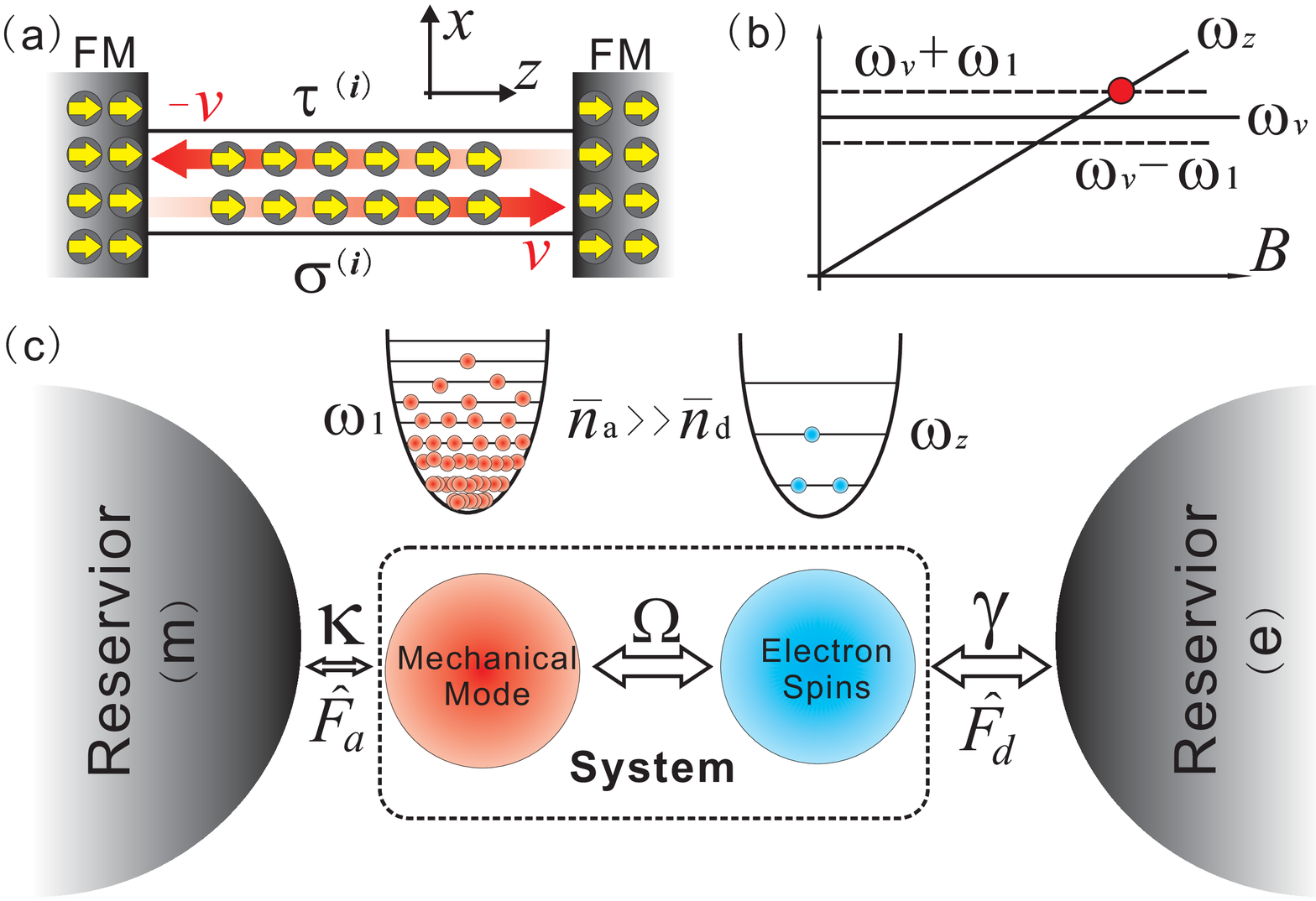}\caption{(a)
Polarized electron spins are injected with velocities $v$ and $-v$
into the NAMR from two ferromagnetic (FM) leads. (b) Controllable
coupling: the collective excitation energy $\omega_{z}$ of electron
spins is linearly dependent on the external magnetic field $B$, and
the frequency $\omega_{v}$ is determined by the electron velocity.
The red point denotes the resonant condition
$\omega_{v}=\omega_{z}-\omega_{1}$ are satisfied. (c) Schematic
illustration of the cooling mechanism. The system contains two
interacting boson modes: the fundamental TNMM (red, left) with
frequency $\omega_{1}$ and the collective spin excitation
mode (blue, right) with frequency $\omega_{z}$.}%
\label{XRef-Figure-41095715}%
\end{figure}

There are three different characteristic frequencies in this system:
the electron Zeeman frequency $\omega_{z}$, the NAMR vibration
frequency $\omega_{1}$, and the the spatial motion induced frequency
$\omega_{v}$. In the interaction picture, the Hamiltonian will
involve coupling terms with different frequencies. In the spirit of
rotating wave approximation, only the terms with the lowest
frequency are retained. Therefore, we can adjust the injecting
velocity and/or the Zeeman energy of the electrons, so that
different types of interaction can be obtained\cite{PRLLiuYX2006}.
For example, if $\omega_{v}\approx\omega_{z}-\omega_{1},$ only two
terms are retained, and we get a JC-type Hamiltonian (see
Fig.\ref{XRef-Figure-41095715}(b)),
$H_{\operatorname{JC}}=g_{v}a^{\dagger}\sigma_{-}\exp[i(\omega_{v}%
t+\phi)]/2+H.c.$.

Having obtained the actively controllable single electron
Hamiltonian, we will study the interactions between the coherently
injected ensembles of electron spins and the TNMM of the NAMR. Let
$N$ spin polarized electrons with velocity $v $ be injected into the
NAMR, and meanwhile, $N$ spin polarized electrons in the opposite
direction with the velocity $-v$. In principle, this model can be
realized by joining the NAMR with two spin polarized electronic
source, for example, two ferromagnetic
leads(Fig.\ref{XRef-Figure-41095715}(a)). Then the Hamiltonian for
this system is
\begin{align}
H_{\operatorname{sys}} &  =\hbar\omega_{1}a^{\dagger}a+\frac{1}{2}\hbar
\omega_{z}(J_{z}+Q_{z})\nonumber\\
&+\tilde{g}_{v}(t)a^{\dagger}(J_{-}-Q_{-})+H.c.,\label{EqTimeDepCoupling}
\end{align}

\noindent where we define the collective spin
operators\cite{PRBSongZ2005}
\begin{align}
J_{z} &  =\sum_{i=1}^{N}\sigma_{z}^{(i)},J_{\pm}=\sum_{i=1}^{N}e^{\mp
i\phi_{i}}\sigma_{\pm}^{(i)},\nonumber\\
Q_{z} &  =\sum_{i=1}^{N}\tau_{z}^{(i)},\ Q_{\pm}=\sum_{i=1}^{N}e^{\mp
i\phi_{i}}\tau_{\pm}^{(i)}.
\end{align}
Here, the Pauli matrices $\sigma_{z,\pm}^{(i)}$ and
$\tau_{z,\pm}^{(i)}$ denote the electron spins with the velocity $v$
and $-v$ respectively, and
$\tilde{g}_{v}(t)=g_{v}\exp(i\omega_{v}t)/2$ is the time dependent
coupling strength.

Through the Holstein-Primakoff transformation $J_{z}=-N+2b^{\dagger
}b,\ \ J_{-}=\sqrt{N}b$ and $Q_{z}=-N+2c^{\dagger}c,\ \
Q_{-}=\sqrt{N}c$, the collective excitation of the ensembles of
electron spins can be characterized by the operators $b$($c$) and
$b^{\dagger}$($c^{\dagger}$), which satisfy the boson commutative
relation in the large N-low excitation limit. Then the interaction
between the collective spin excitation and NAMR is modeled by an
interacting two-mode boson system. In the interaction picture,
\begin{equation}
H_{I}=\hbar\Omega(a^{\dagger}de^{-i\Delta t}%
+H.c.),\label{XRef-Equation-41095042}%
\end{equation}

\noindent where $d=(b-c)/\sqrt{2}$ is the boson operator associated
with the collective spin excitations, $\hbar\Omega=\sqrt{N/2}g_{v}$
is the Rabi frequency, and $\Delta=\omega_{z}-\omega_{1}-\omega_{v}$
is the detuning.

\emph{Cooling mechanism}.-Next, we study the active cooling
mechanism based on the dynamical thermalization of the coupled
system of the TNMM and ensembles of electron spins
(Fig.\ref{XRef-Figure-41095715}(c)). To describe a practical
situation, we assume the TNMM and the ensembles of electron spins to
interact with two separated thermal reservoirs. The time evolution
of the system operators $a$ and $d$ are governed by the
Heisenberg-Langevin equantions\cite{bookScullyQuantumOptics1997}
\begin{gather}
\dot{a}=-i\Omega d e^{-i\Delta t}-\frac{\kappa}{2}a+\hat{F}_{a}%
(t)\label{XRef-Equation-41615726}\\
\dot{d}=-i\Omega a e^{i\Delta t}-\frac{\gamma}{2}d+\hat{F}_{d}%
(t)\label{XRef-Equation-4161580}%
\end{gather}

\noindent where $\hat{F}_{a}(t)$ and $\hat{F}_{d}(t)$ are noise
operators due to the reservoirs. We assume the two reservoirs are
statistically independent, i.e.
$\langle\hat{F}_{a}^{\dagger}(t)\hat{F}_{d}(t^{\prime})\rangle=0$.
The correlations beteen $\hat{F}_{a}(t)$ and
$\hat{F}_{a}^{\dagger}(t)$
\begin{align}
\langle\hat{F}_{a}^{\dagger}(t)\hat{F}_{a}(t^{\prime})\rangle &  =\kappa
\bar{n}_{a}\delta(t-t^{\prime}),\nonumber\\
\langle\hat{F}_{a}(t)\hat{F}_{a}^{\dagger}(t^{\prime})\rangle &
=\kappa (\bar{n}_{a}+1)\delta(t-t^{\prime})
\end{align}
describe the quantum fluctuations with $\kappa$ being the damping constant and
$\bar{n}_{a}=[\exp(\hbar\omega_{1}/k_{B}T)-1]^{-1}$ being the thermal
occupation number of TNMM. The noise operators $\hat{F}_{d}(t)$ and $\hat
{F}_{d}^{\dagger}(t)$ follow similar relations with damping constant $\gamma$
and thermal occupation number $\bar{n}_{d}=[\exp(\hbar\omega_{z}%
/k_{B}T)-1]^{-1}$.

\begin{figure}[ptb]
\includegraphics[bb=16 15 315 217,angle=0, width=7 cm, clip]{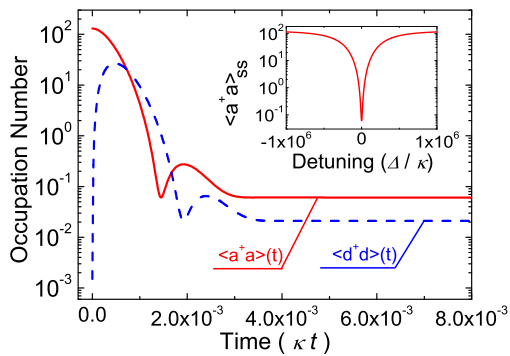}
\caption{Time evolution of the occupation numbers $\langle a^{\dagger}%
a\rangle(t)$ and $\langle d^{\dagger}d\rangle(t)$. The occupation numbers of
the two boson modes reach steady state values when $t\rightarrow+\infty$.
Inset: the steady state occupation number $\langle a^{\dagger}a\rangle
_{\operatorname{ss}}$ as a function of the detuning $\Delta$. Parameters used
in the calculation are described in text.}%
\label{XRef-Figure-41010246}%
\end{figure}

It follows from Eqs.(\ref{XRef-Equation-41615726}) and
(\ref{XRef-Equation-4161580}) that the motions of bilinear quantities
${\mathcal{A}}=\langle a^{\dagger}a+aa^{\dagger}\rangle$, ${\mathcal{D}%
}=\langle d^{\dagger}d+dd^{\dagger}\rangle$, and ${\mathcal{C}}=\langle
ad^{\dagger}+d^{\dagger}a\rangle$ are
\begin{gather}
\dot{{\mathcal{A}}}=i\Omega\tilde{{\mathcal{C}}}-i\Omega\tilde{{\mathcal{C}}%
}^{\ast}-\kappa{\mathcal{A}}+\kappa\left(  2\bar{n}_{a}+1\right), \\
\dot{{\mathcal{D}}}=-i\Omega\tilde{{\mathcal{C}}}+i\Omega\tilde{{\mathcal{C}}%
}^{\ast}-\gamma{\mathcal{D}}+\gamma\left(  2\bar{n}_{d}+1\right), \\
\dot{\tilde{{\mathcal{C}}}}=i\Omega{\mathcal{A}}-i\Omega{\mathcal{D}}-\frac
{1}{2}(\kappa+\gamma)\tilde{{\mathcal{C}}},%
\end{gather}

\noindent where $\tilde{{\mathcal{C}}}={\mathcal{C}}\exp(i\Delta
t)$. The above set of equations determines the time evolution of the
occupation numbers (Fig.\ref{XRef-Figure-41010246}). At steady
state, the mean occupation number of the TNMM is
\begin{equation}
\langle a^{\dagger}a\rangle_{\operatorname{ss}}=\bar{n}_{a}-\gamma\left(
\bar{n}_{a}-\bar{n}_{d}\right)  {\mathcal{L}}(\kappa,\gamma,\Delta
),\label{XRef-Equation-321161339}%
\end{equation}

\noindent where ${\mathcal{L}}( \kappa,\gamma,\Delta) =4\Omega^{2}(
\kappa+\gamma) /\big[\kappa\gamma[ (\kappa+\gamma)^{2}(4\Omega^{2}%
/\kappa\gamma+1)+\Delta^{2}]\big] $ is a Lorentz line shape function.

The first term in Eq. (\ref{XRef-Equation-321161339}) represents the
initial thermal occupation number of the TNMM in equilibrium with
the thermal reservoir without injection electron spins. And the
second term, which is proportional to the coupling strength
$\Omega^{2}$, is due to the interaction with the spin polarized
electrons. It is clear that, if $\bar{n}_{a}>\bar {n}_{d}$, the mean
occupation number $\langle a^{\dagger}a\rangle _{\operatorname{ss}}$
will be reduced to a value lower than $\bar{n}_{a}$, or in other
words, the TNMM is cooled down. Furthermore, a large cooling
efficiency demands the condition $\bar{n}_{a}>>\bar{n}_{d}$. It is
equivalence to require that (i) the electron Zeeman energy
$\omega_{z}$ is much larger than the TNMM frequency $\omega_{1}$,
i.e. $\omega_{z}>>\omega_{1}$. This is the first necessary condition
for a high cooling efficiency. Under this condition, the mean
occupation number at resonance ($\Delta=0$) can be rewritten as the
weighted average of the thermal occupation numbers $\bar {n}_{a}$
and $\bar{n}_{d}$:
\begin{equation}
\langle a^{\dagger}a\rangle_{\operatorname{ss}}=\left(
1-f_{1}f_{2}\right) \bar{n}_{a}+f_{1}f_{2}\bar{n}_{d}
\end{equation}

\noindent with the two ratios $f_{1}=4\Omega^{2}/(\kappa\gamma+4\Omega^{2})$
and $f_{2}=\gamma/(\gamma+\kappa)$. Thus, we find that the mean occupation
number $\langle a^{\dagger}a\rangle_{\operatorname{ss}}$ has a neglectably
small lower bound $\bar{n}_{d}$, and a significant reduction of $\langle
a^{\dagger}a\rangle_{\operatorname{ss}}$ with respect to the initial thermal
equilibrium value $\bar{n}_{a}$ needs the following two additional
requirements: (ii) the coupling between the two modes $a$ and $d$ are strong
enough compared with the dampings, i.e. $4\Omega^{2}>>\kappa\gamma$, and (iii)
the decay rate of the collective electron spin excitations is much larger than
the one of the TNMM, i.e. $\gamma>>\kappa$.

To explain the physical mechanism beneath the above protocol, we
suppose that the two boson modes are initially in equilibrium with
their reservoirs of the same temperature $T$. Under the condition
(i), the initial occupation numbers $\bar{n}_{a}>>\bar{n}_{d}$ (see
Fig. \ref{XRef-Figure-41095715}(c)). The condition (ii) ensures the
efficient population transfer between the two boson modes. For
$t>0$, the occupation number of the TNMM (collective excitation of
electron spins) will decrease (increase) due to the coupling (Fig.
\ref{XRef-Figure-41010246}). Simultaneously, the reservoirs tend to
maintain the thermal occupation numbers $\bar{n}_{a}$ and
$\bar{n}_{d}$. In other words, the occupation number of the TNMM
(collective excitation electron spins) will gain from (decay to) the
reservoir with the rate $\kappa$ ($\gamma$). The condition (iii)
defines two different time scales of the thermalization processes.
The much faster decay rate $\gamma$ than $\kappa$ guarantees that
the net effect of the dynamical thermalization is reduction of the
occupation number.

Here, we emphasize the importance of the time dependence of coupling
$\tilde{g}_{v}(t)$(see Eq. \ref{EqTimeDepCoupling}) induced by the
spatial motion of spins. Generally speaking, the frequency mismatch
of two interacting boson modes will prohibit any efficient
couplings. In other words, the requirement (i) would block off the
occupation transfer between the two boson modes if their coupling
were time independent. Fortunately, in our model, the spatial motion
provides a third frequency $\omega_{v}$ to compensate the large
frequency difference between $\omega_{1}$ and $\omega_{z}$. Thus,
the generic resonant condition $\omega_{z}=\omega_{1}$ is modified
to $\omega_{v}=\omega_{z}-\omega_{1}$. In this way, the spatial
motion ensures an efficient coupling under the frequency mismatch
condition (i).

Finally, we discuss the experimental feasibility of our cooling mechanism. For
a typical NAMR with $l=2\mu m$, $d=100\operatorname{nm}$, and
$h=80\operatorname{nm}$, and the TNMM with frequency $\omega_{1}%
=300\operatorname{MHz}$, the dimensionless constant ${\mathcal{E}}%
\approx10^{-8}$ according to its definitions. With the parameters of
GaAs material, mass density $\rho=5.3g/\operatorname{cm}^{3}$,
effective electron mass $m=0.067m_{e}$, the spin-orbit interaction
constant $\alpha/\hbar
=4\times10^{5}\operatorname{m}/\operatorname{s}$, and the velocity
of injected electrons
$v_{F}=1.6\times10^{5}\operatorname{m}/\operatorname{s}$, which
corresponds to Fermi energy $E_{F}=5\operatorname{meV}$, we estimate
the coupling strength $\Omega=7\operatorname{MHz}$ for $N=400$. The
decay rate $\gamma$ for the collective excitation of electron spins
is determined by the electron spin coherence time
$T_{1}\sim50\operatorname{ns}$, which implies
$\gamma\sim1/T_{1}=20\operatorname{MHz}$. For the quality factor
$Q=10^{5}$ of the NAMR , and thus
$\kappa=\omega_{1}/Q=3\operatorname{kHz}$, we find the occupation
number could reduced from $\bar{n}_{a}=130$ without coupling down to
$\langle a^{\dagger}a\rangle_{\operatorname{ss}}=0.06<<1$ at a
sub-Kelvin temperature $T=300\operatorname{mK}$.

\emph{Conclusion}.-We investigate a spin-TNMM coupling model for
cooling the TNMM through dynamical thermalization. We found that the
strain induced spin-orbit interactions for conduction band electrons
in semiconductor materials result in a microscopic coherent coupling
between the electron spins and the quantized TNMM. With this
discovery, we present an experimentally accessible cooling method
for the TNMM by injecting ensembles of electron spins. The TNMM can
be sufficiently cooled down to be capable of exhibiting various
quantum phenomena.

This work is supported by the NSFC with grant Nos. 90203018,
10474104, 10574077, and 60433050, and NFRPC with Nos. 2006CB921206,
2006CB0L0601, 2006AA06Z104 and 2005CB724508.

\appendix

\end{document}